\documentclass[review]{elsarticle}
\usepackage{amssymb}
\usepackage{graphicx}
\usepackage{marginnote}
\usepackage{url}
\usepackage{color}

\begin{document}
\suppressfloats

\title{Ideal wet two-dimensional foams and emulsions with finite contact angle}
\author{S.J. Cox$^{(a)}$, A.M. Kraynik$^{(b,c)}$, D. Weaire$^{(c)}$ and S. Hutzler$^{(c)}$}
\address{
(a) Department of Mathematics,
Aberystwyth University, Aberystwyth SY23 3BZ, UK \\
(b) Sandia National Laboratories (Retired) \\
(c) School of Physics, Trinity College Dublin, The University of Dublin, 
Ireland 
}

\begin{abstract}
We present simulations that show that an ideal two-dimensional foam  with a finite contact angle develops an inhomogeneity for
high liquid fraction $\phi$. In liquid-liquid emulsions this inhomogeneity is known as flocculation.
In the case of an ordered foam this requires
a perturbation, but in a disordered foam inhomogeneity grows steadily and
spontaneously with $\phi$, as demonstrated in our
simulations performed with the Surface Evolver.
\end{abstract}
\maketitle

\section{Introduction}
\begin{figure}
\begin{center}
(a)\includegraphics[width=0.4\textwidth]{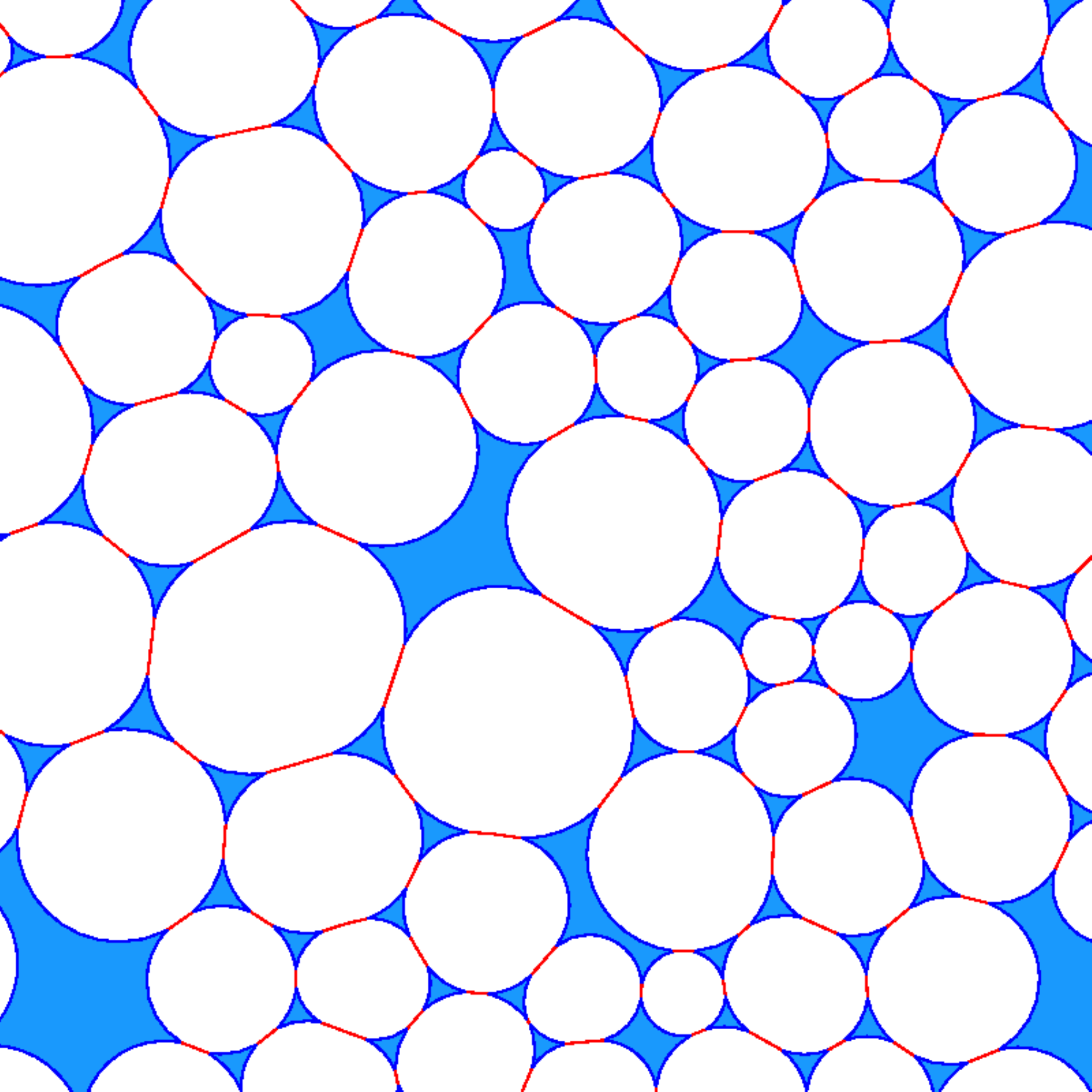} \hspace{0.5cm}
(b)\includegraphics[width=0.4\textwidth]{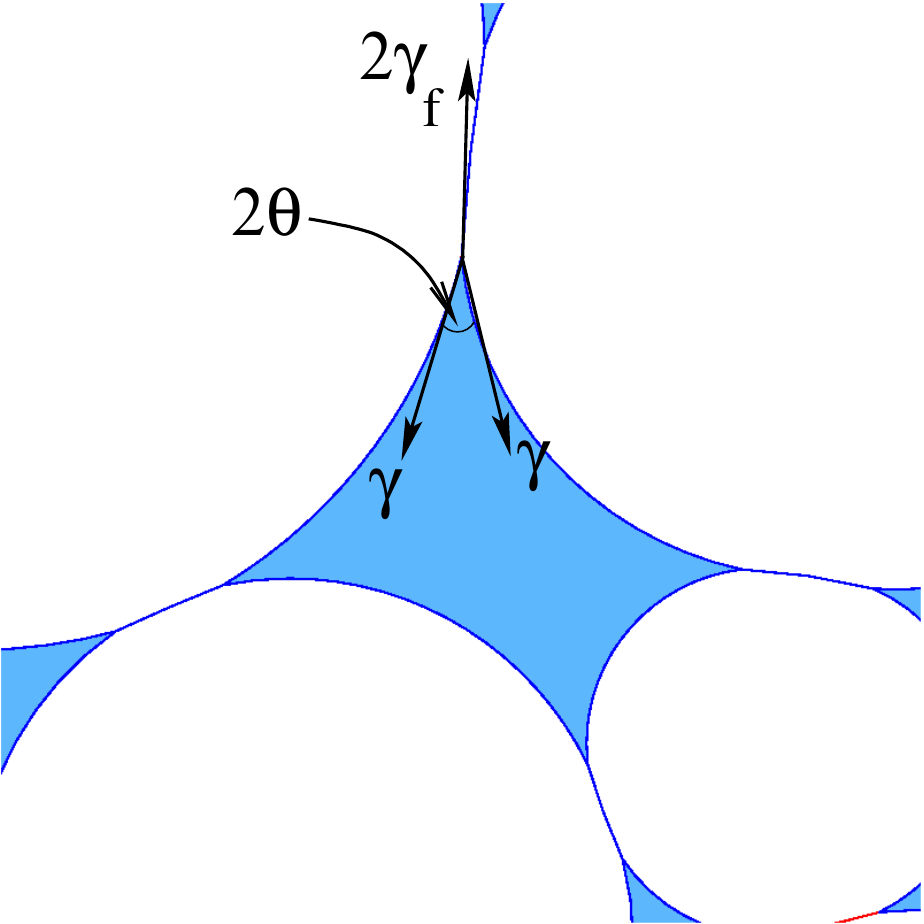}
\end{center}
\caption{
(a) Typical simulation of a
two-dimensional foam with liquid fraction $\phi = 0.1$ and contact angle
$\theta = 9.5^\circ$.
The image shows part of an equilibrated foam of 1500 bubbles.
(b) Close-up of a four-sided Plateau border, showing the contact angle
$\theta$. The vectors represent the forces, due to interfacial tension, acting on the
point of contact.
}
\label{f:se_sample}
\end{figure}

In emulsions, the term flocculation refers to the (spontaneous) clustering of droplets,
leading to the formation of density inhomogeneities. 
Here we describe the onset of flocculation in computer simulations of
two-dimensional (2D) liquid foams. This only occurs in systems where the liquid-gas
interfaces meet at a finite contact angle $\theta$, as illustrated in figure~\ref{f:se_sample}(b).
Previous simulations of 2D foams with finite liquid fraction $\phi$ have taken this contact angle to be zero \cite{boltonweaire92} or, where that was not feasible for numerical reasons, as small as possible \cite{jing15}.
We treat both liquid and gas as incompressible, and so the results apply equally to emulsions.

Two-dimensional foams have properties that are broadly similar to their
three-dimensional counterparts, but are much simpler to analyse.
Their study has early antecedents~\cite{Smith52}, and continues to
be of interest today. The usual theoretical model is entirely
two-dimensional, while the third dimension may be significant in relevant
experimental systems, such as a foam trapped between two plates~\cite{Smith52,CoxJaniaud08}.

A simulated example of an ideal 2D foam with finite liquid fraction $\phi$ and finite contact angle $\theta$ is shown in figure~\ref{f:se_sample}(a). 
It was produced by the  method described in \ref{a:se}. 
The gas bubbles are surrounded by a network of smoothly-curved thin films connecting the liquid-filled Plateau borders, which each have three or more sides. There are therefore two types of interface: liquid-gas interfaces around each Plateau border, and gas-liquid-gas interfaces forming the bubble-bubble contacts. On each of the interfaces the Laplace-Young law
relates the product of interfacial tension and curvature to 
the pressure difference across the interface~\cite{HutzlerBook,Cantat}, 
and consequently liquid films are represented by arcs of circles that meet at the vertices of Plateau borders.
The films are considered to be infinitesimally thin, so all of the liquid in the foam
is considered to be contained in the Plateau borders.
 
A finite contact angle implies that the interfacial tension associated with the bubble-bubble interfaces is less than twice that associated
with the Plateau borders (see figure \ref{f:se_sample}(b)). 
Note that throughout this paper ``interfacial tension'' is used for what is
really a {\em line} tension (or energy per unit line length) in such an idealised 
2D model.
The contact angle $\theta$ is given
by 
\begin{equation}
\theta = \cos^{-1} \left(\frac{\gamma_f}{\gamma}\right),
\label{e:contact_angle}
\end{equation}
where $\gamma_f$ is the interfacial tension of each side of a liquid film;
this is equal to or smaller than the
(bulk) interfacial tension $\gamma$ associated with the gas-liquid
interfaces of a Plateau border~\cite{Princen1979}.
The surface energy $E(\phi,\theta)$ of the foam is the sum of the lengths of all interfaces, 
multiplied by their appropriate interfacial tension.

\begin{figure}
\begin{center}
\includegraphics[width=0.8\textwidth]{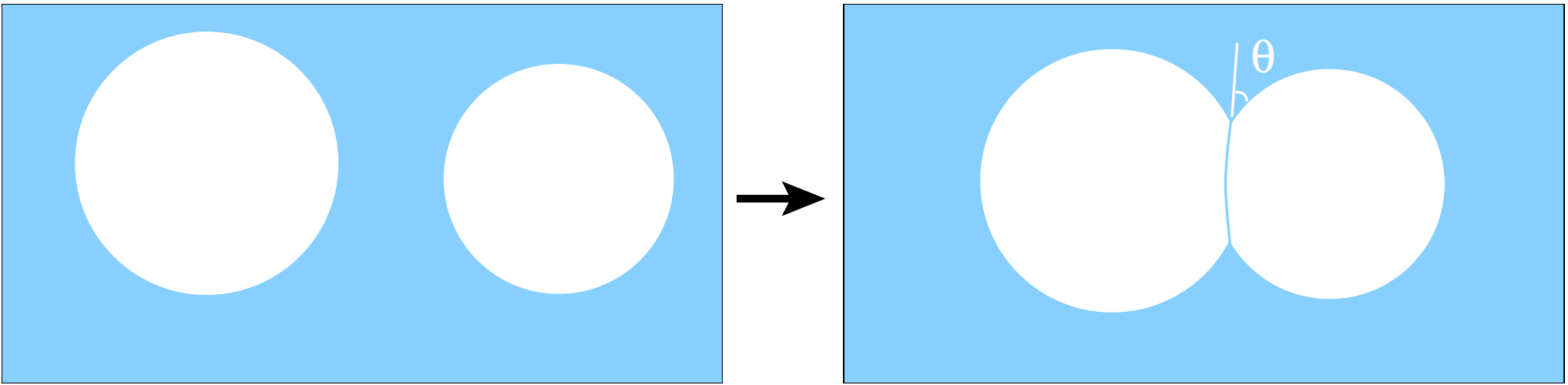}
\end{center}
\caption{The (line) energy of two
isolated circular bubbles is reduced when they share a common interface, if
the contact angle $\theta$ is finite.
}
\label{f:clumping}
\end{figure}

For bubbles immersed in a liquid, the presence of finite contact angles
entails net attractive forces between them (see figure \ref{f:clumping})
when they are only slightly compressed together. This is similar to the attraction between droplets, 
when considering emulsions~\cite{Langevin2015}.

In the present paper we address some basic consequences of introducing
finite contact angles into the standard model of 2D foams,  by analysing
simulations carried out with the Surface Evolver software
of Ken Brakke~\cite{Brakke92}. We will see that finite
contact angles, even if apparently very small, can have large effects.

Henry Princen introduced the concept of a contact angle 
between a thin liquid film and its adjacent Plateau border~\cite{Princen1965} 
and he established its physical significance in foams and emulsions by analysing 
2D \mbox{{\em ordered}} (hexagonal) monodisperse structures~\cite{Princen1979}, 
which admit analytical solutions. 
The Surface Evolver and current computational resources 
enable the simulation of {\em disordered} foams, 
which is the usual practical case of interest.
Princen's work was stimulated by his own measurements of contact angles
for soap films in contact with bulk solution~\cite{princen1968contact}.
He found that finite contact angles up to $17^\circ$ could be achieved in
surfactant
(SDS) solutions at sufficiently high
concentrations of added electrolytes.

We shall begin by recapitulating Princen's 2D model, which gave a number of exact results. 
This turns out to be instructive when discussing our own 
results from simulations of disordered 
polydisperse foams.

\section{Ordered hexagonal foams}

Even the apparently trivial case of an ordered 2D foam proves to present
some challenges to detailed understanding, so we shall examine it carefully.

\begin{figure*}
\begin{center}
(a)\includegraphics[width=0.3\textwidth]{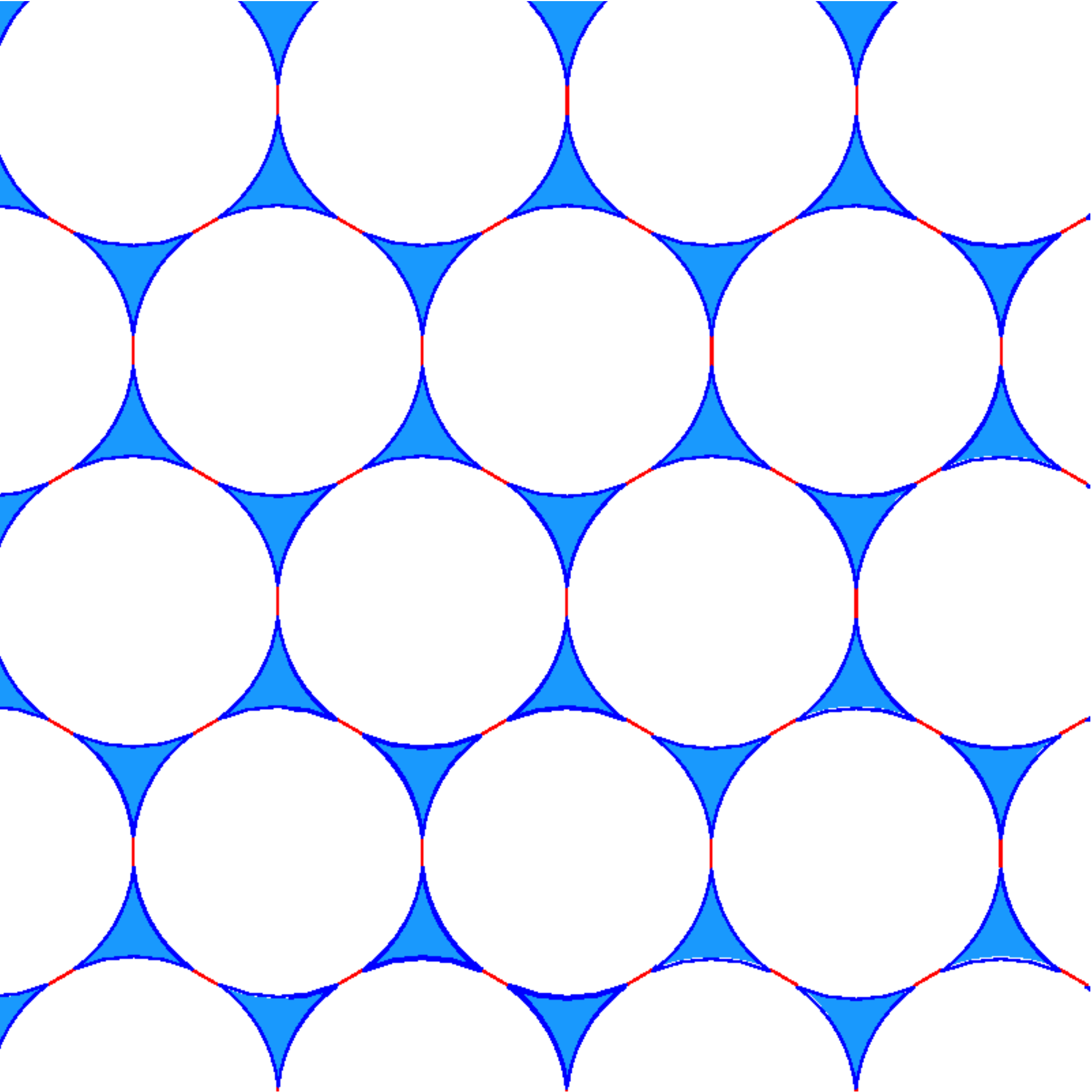} \hspace{1cm}
(b)\includegraphics[width=0.3\textwidth]{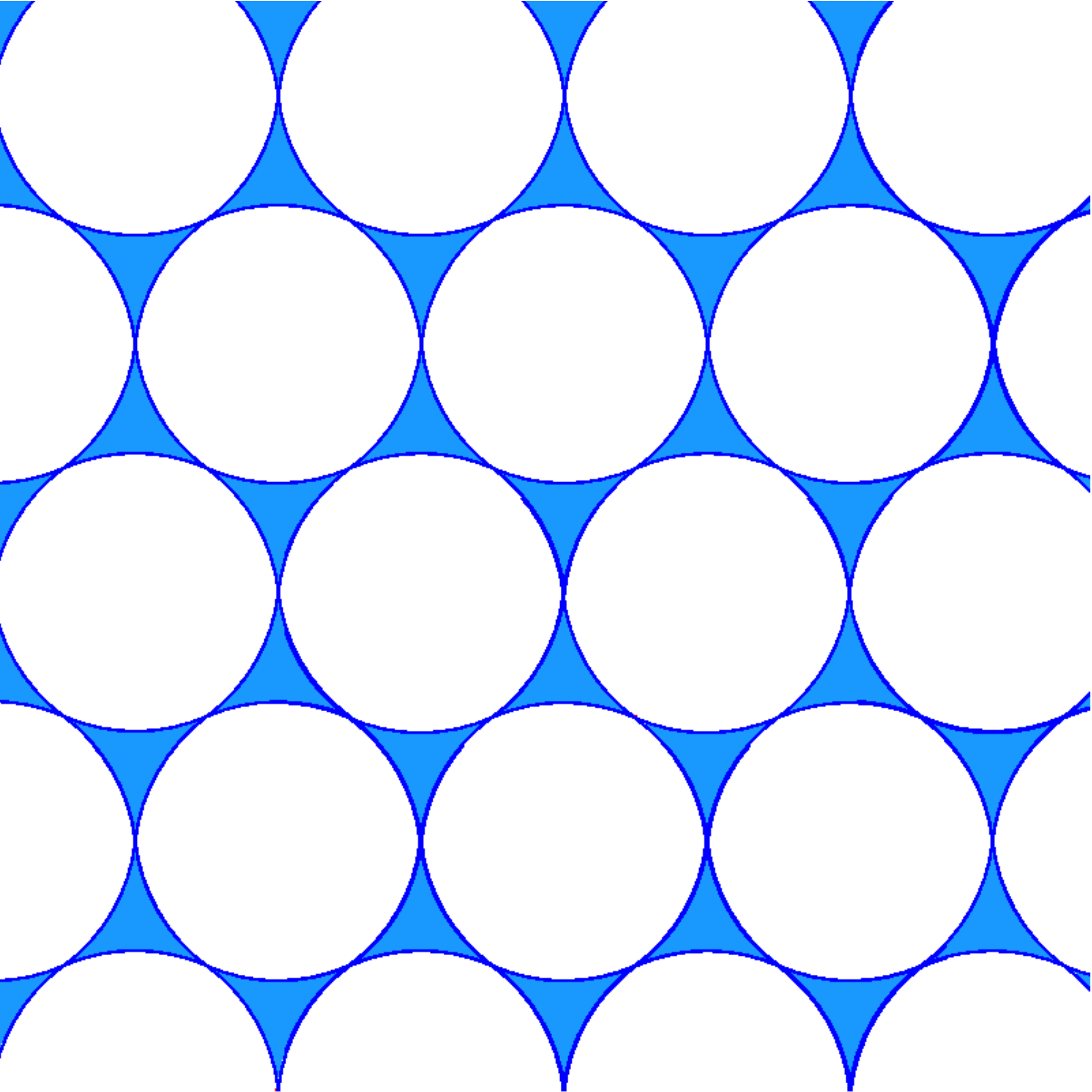}
\hspace{1cm}
\end{center}
\caption{
Examples of monodisperse hexagonal foams for contact angle $\theta = 6.26^\circ$ and 
different values of liquid fraction.
(a)  Liquid fraction $\phi_0^\theta = 0.084$, where the energy is minimal (state of zero
compression).
(b)  Liquid fraction $\phi_m^\theta = 0.128$ (eqn.~(\ref{e:phi_0_theta})), the maximum value of liquid
fraction at which the bubbles still remain
in contact (eqn.~(\ref{e:phi_m_theta})). 
}
\label{f:phimax}
\end{figure*}

Monodisperse 2D bubbles are arranged in equilibrium 
on an hexagonal lattice (figure
\ref{f:phimax}). Princen~\cite{Princen1979}
showed that the liquid fraction $\phi_0^{\theta}$ of such a packing {\em at
zero compression}, which corresponds to zero {\em osmotic 
pressure}~\cite{Princen1979,Princen86osmotic,Hutzler1995osmotic}) 
and minimal surface energy~\cite{Princen1979}, is given by 
\begin{equation}
\phi_0^{\theta} = 1-\frac{\pi - 6 \theta+
3\sin(2\theta)}{2\sqrt{3}\cos^2\theta}.
\label{e:phi_0_theta}
\end{equation}
In the case of a zero contact angle ($\theta = 0$),  $\phi_0^{\theta}$
reduces to the familiar value 
\begin{equation}
\phi_0:= \phi_0^{\theta=0} = 1-\pi/(2\sqrt{3}) \simeq 0.093,
\end{equation}
which is the liquid fraction of an hexagonal close packing of circular bubbles; 
it is known as the {\em wet limit} of an ordered monodisperse foam with zero contact angle.
However, the bubbles are never circular when the contact angle is finite.
Figure \ref{f:phimax} shows two examples of an ordered foam with finite contact angle, 
for different values of the liquid fraction.

Princen calculated
the work {\em per unit area}, 
$\Delta W^\theta$, required to compress 
(by removing liquid)
a foam from the liquid fraction
$\phi_0^\theta$ to any given $\phi$~\cite{Princen1979}.
The bubble area is written in terms of the radius $R$ of an
undeformed circular bubble of the same area, 
resulting in\footnote{Princen~\cite{Princen1979} describes his calculation in terms of the deformation
of columns of hypothetical cylindrical emulsion drops. We have re-written his
expression in terms of liquid fraction $\phi$, rather than gas fraction $(1-\phi)$, and
corrected one misprint (missing superscript $\theta$ in his equation
(35)).}
\begin{equation}
\frac{\Delta W^\theta}{\gamma/R} = 2 (1-\phi_0^\theta)
\cos\theta\left[\frac{1}{(1-\phi)^{1/2}(1-\phi_0)^{1/2}} -
\left( \frac{\phi_0^\theta}{1-\phi_0}
\right)^{1/2}\left(\frac{\phi}{1-\phi} \right)^{1/2} - \left(
\frac{1-\phi_0^\theta}{1-\phi_0} \right)^{1/2} \right],
\label{e:princen-work}
\end{equation}
where the normalizing factor $\gamma/R$ is often called the
Laplace pressure.

In the following we will consider the 
energy $E(\phi,\theta)$ {\em per bubble} 
as a function of liquid fraction $\phi$ and contact angle $\theta$.
Using eqn.~(\ref{e:princen-work}) and
$\Delta W^\theta=  \displaystyle{\frac{1-\phi_0^\theta} {R^2 \pi}}
(E(\phi,\theta)-E(\phi_0^\theta,\theta))$, from the work-energy theorem,
we obtain
\begin{equation}
\frac{E(\phi,\theta)}{2\pi R \gamma \cos\theta} =
\frac{1 - \sqrt{\phi_0^\theta \phi}} {\sqrt{(1-\phi)(1-\phi_0)}}.
\label{e:e_phi_gamma}
\end{equation}
In the
dry limit ($\phi=0$, i.e., close-packed hexagons), 
this reduces to 
$E(0,\theta)/(2\pi R \gamma\cos\theta)= (1-\phi_0)^{-1/2}$.

\begin{figure*}
\begin{center}
\includegraphics[width=0.7\textwidth]{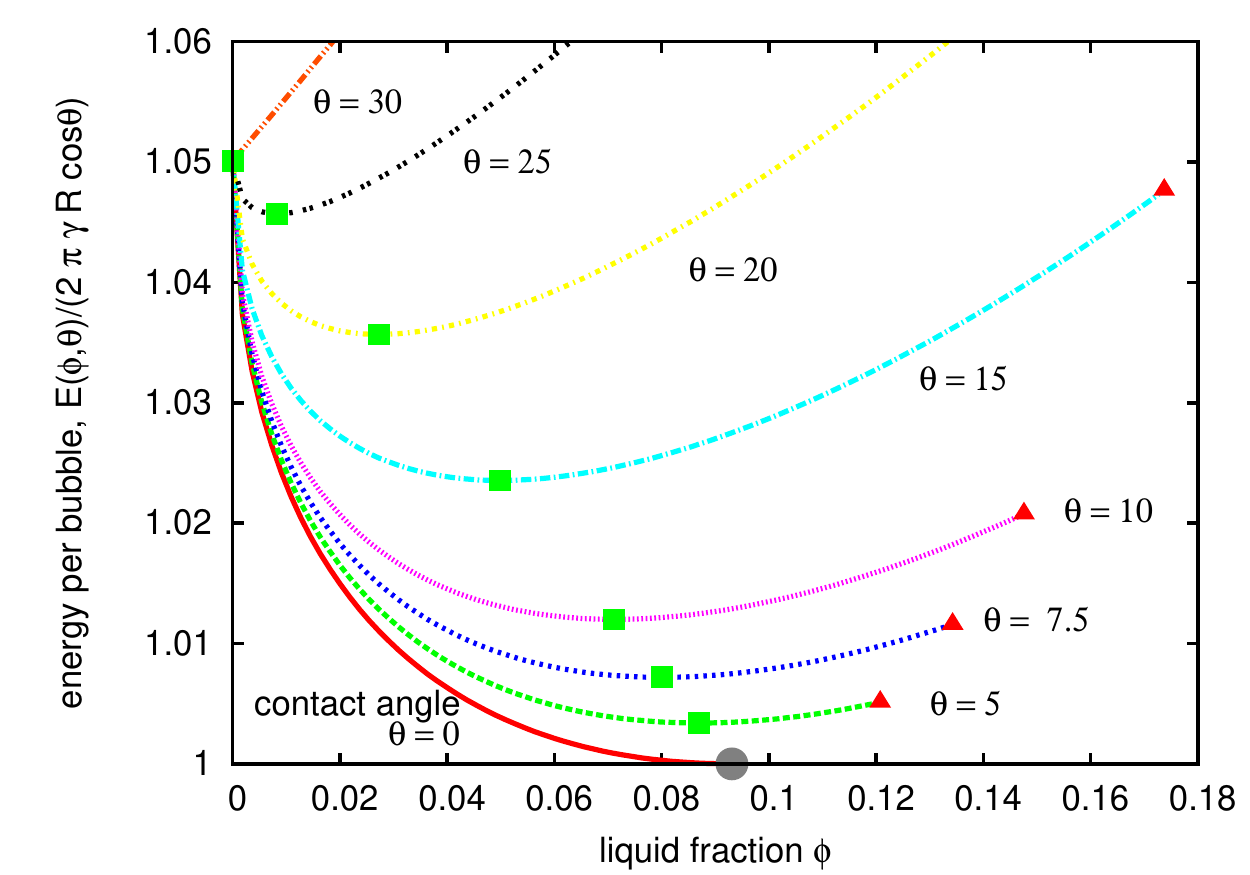}
\end{center}
\caption{
Variation of 
energy per bubble $E(\phi,\theta)/(2 \pi R \gamma \cos\theta)$ (eqn.~(\ref{e:e_phi_gamma}))
of an ordered hexagonal 2D foam with liquid fraction $\phi$, for a
range of different contact angles $\theta$. ($\gamma$: bulk interfacial
tension, $R$: radius of an undeformed (circular) bubble).
Symbols correspond to the
critical liquid fractions $\phi_0^\theta$ (green
squares), $\phi_m^\theta$ (red triangles), 
and $\phi_0$ (grey disc), defined in the text.
}
\label{f:energy_finite_contact}
\end{figure*}

The energy $E(\phi,\theta)$ in eqn.~(\ref{e:e_phi_gamma}) 
may be characterised by {\em two} 
different critical values of the liquid fraction, 
as shown in Figure \ref{f:energy_finite_contact}.
These are
\begin{itemize}
\item the liquid fraction $\phi_0^{\theta}$ (eqn.~(\ref{e:phi_0_theta})) at which
the energy has a minimum;
\item the maximum value of liquid fraction
$\phi_{m}^{\theta}$
at which the bubbles remain in contact, given by 
\begin{equation}
\phi_{m}^{\theta} = 
\frac{\sqrt{3}}{8 \sin^2\frac{\varphi}{2}} 
\left[
2 \sin\frac{\varphi}{2}
\left(
\frac{1}{\sqrt3} \sin\frac{\varphi}{2} + \cos\frac{\varphi}{2}
\right)
- \varphi
\right],
\label{e:phi_m_theta}
\end{equation}
with $\varphi = \pi/3 - 2\theta$. 
This is the hypothetical wet limit of an ordered monodisperse foam with {\em finite} contact angle.
\end{itemize}

\begin{figure*}
\begin{center}
\includegraphics[width=0.7\textwidth]{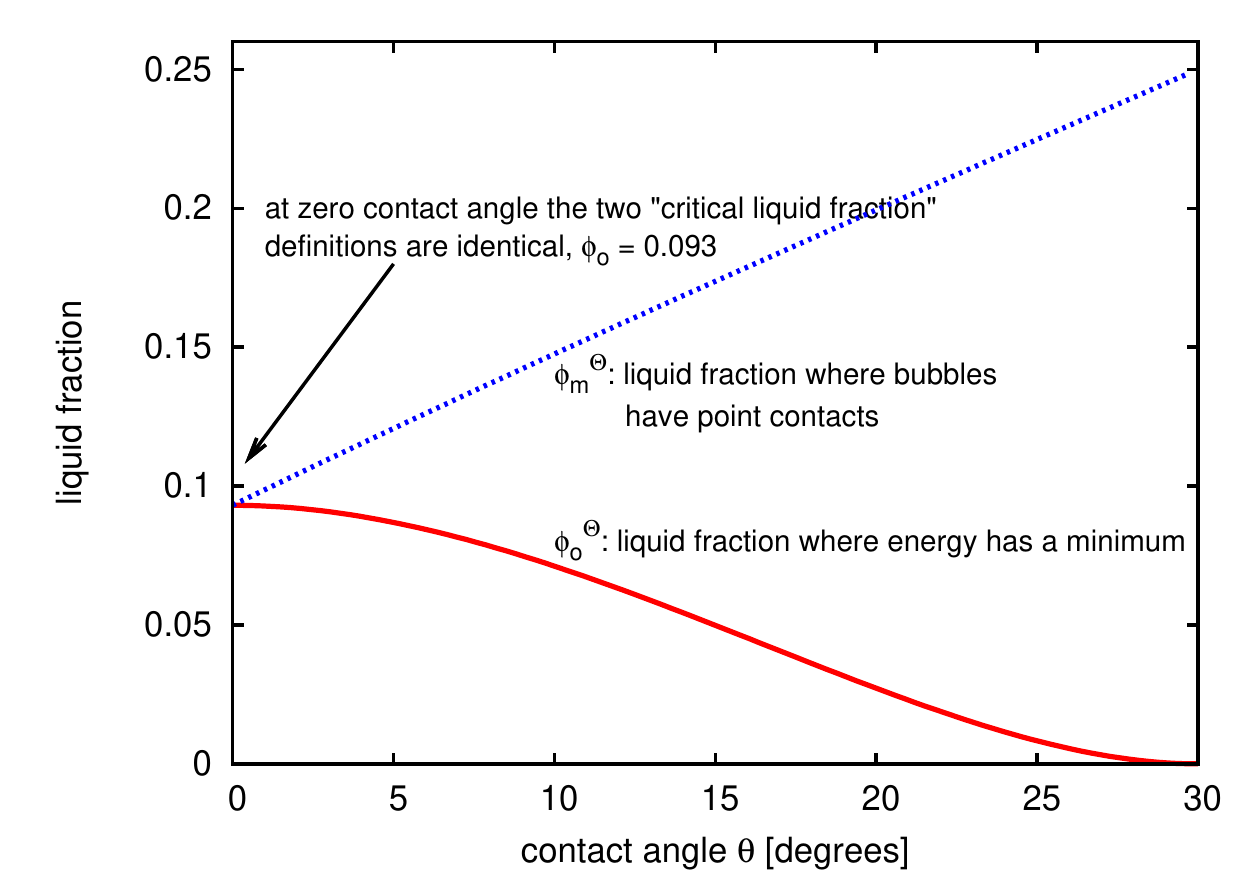}
\end{center}
\caption{
Variation of the critical liquid fractions $\phi_0^{\theta}$
(eqn.~(\ref{e:phi_0_theta})), and
$\phi_m^{\theta}$ (eqn.~(\ref{e:phi_m_theta}))
as a function of contact
angle $\theta$ in an ordered hexagonal foam.
}
\label{f:phi_finite_contact}
\end{figure*}

Figure \ref{f:phi_finite_contact} displays the
variation of $\phi_0^{\theta}$ and $\phi_m^{\theta}$  with 
contact angle $\theta$.
Note that $\theta=0$ is a special case because the critical values coincide,  $\phi_0^0=
\phi_m^0 = \phi_0 = 1-\pi/(2\sqrt{3}) \simeq 0.093$.
Figure \ref{f:phimax} shows an ordered foam at each of the critical liquid fractions $\phi_m^\theta$ and
$\phi_0^\theta$, for contact angle $\theta = 6.26^\circ$.

When the contact angle is finite, the non-circular bubbles at $\phi_{m}^{\theta}$ 
contact their neighbours at a point. However, this situation differs from the conventional wet limit for zero contact angle, as we explore below.

\begin{figure*}
\begin{center}
\includegraphics[width=0.7\textwidth]{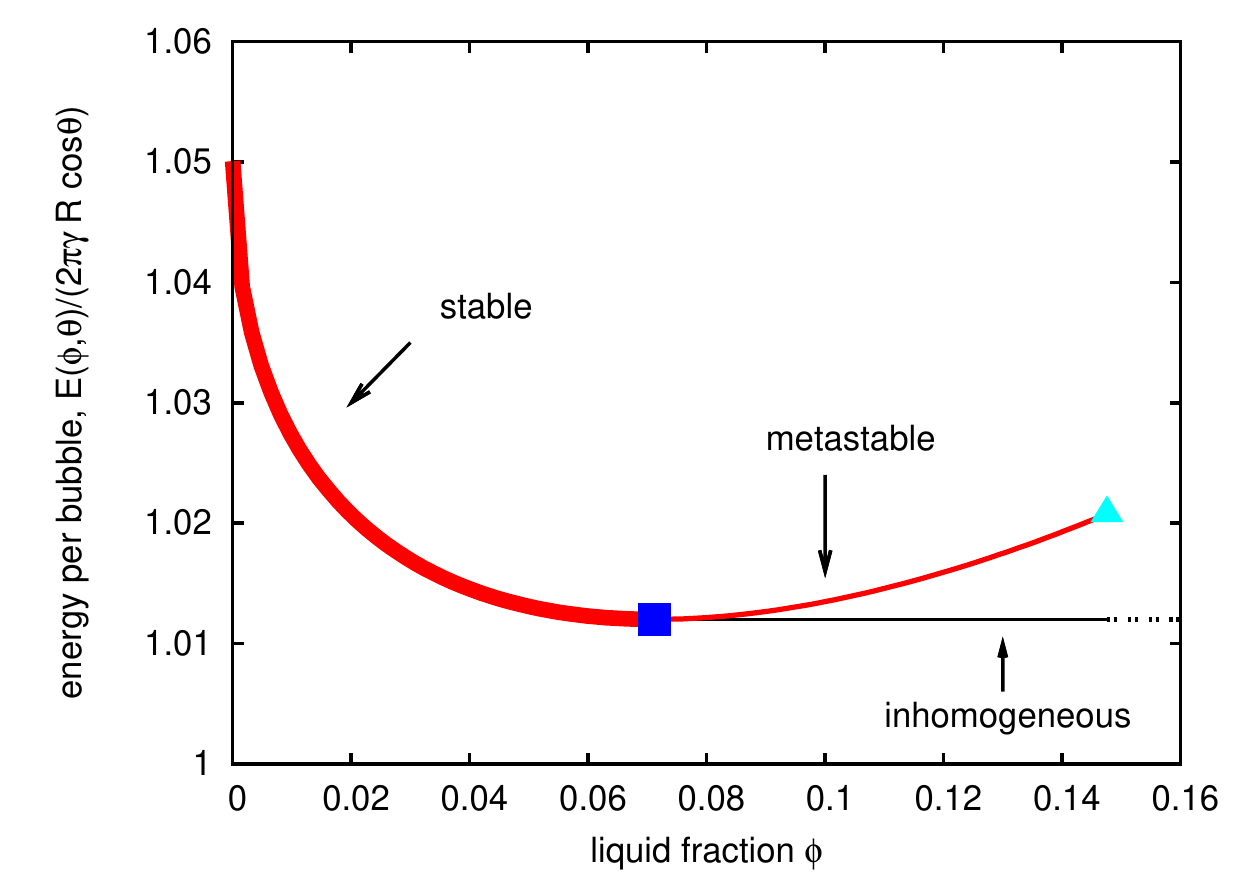}
\end{center}
\caption{
Variation of energy $E(\phi,\theta)$ of an initially ordered hexagonal foam
(for contact angle $\theta=10^\circ$). 
For values of liquid fraction exceeding $\phi_0^\theta$ (marked by a blue
square), the hexagonal
bubble arrangement is metastable, since alternative inhomogeneous structures
exist with lower energy. (Figure  \ref{f:honey_phase_separated}(a) shows such
an example, obtained from simulation.)
The light blue triangle at critical liquid fraction $\phi_m^\theta$
corresponds to the point where bubbles in an hexagonal arrangement no longer
touch.
}
\label{f:t-line}
\end{figure*}

All points on the curves for energy as a function of liquid fraction, 
shown in figure \ref{f:energy_finite_contact}, correspond to  (possibly metastable)
equilibrium structures. 
Since the liquid fraction $\phi_0^{\theta}$ corresponds to the {\em minimum}
energy, any homogeneous structure for which $\phi > \phi_0^{\theta}$ must be
metastable, at least for an infinite sample. 
This is because an {\em inhomogeneous} structure can be defined, 
with constant energy (in the limit of infinite sample size) close to $E(\phi_0^\theta)$. 
This scenario is illustrated in figure \ref{f:t-line}, where we have
represented the energy of the inhomogeneous structure
by a horizontal line beyond $\phi_0^{\theta}$.

\begin{figure*}[]
\begin{center}
(a) \includegraphics[width=0.4\textwidth]{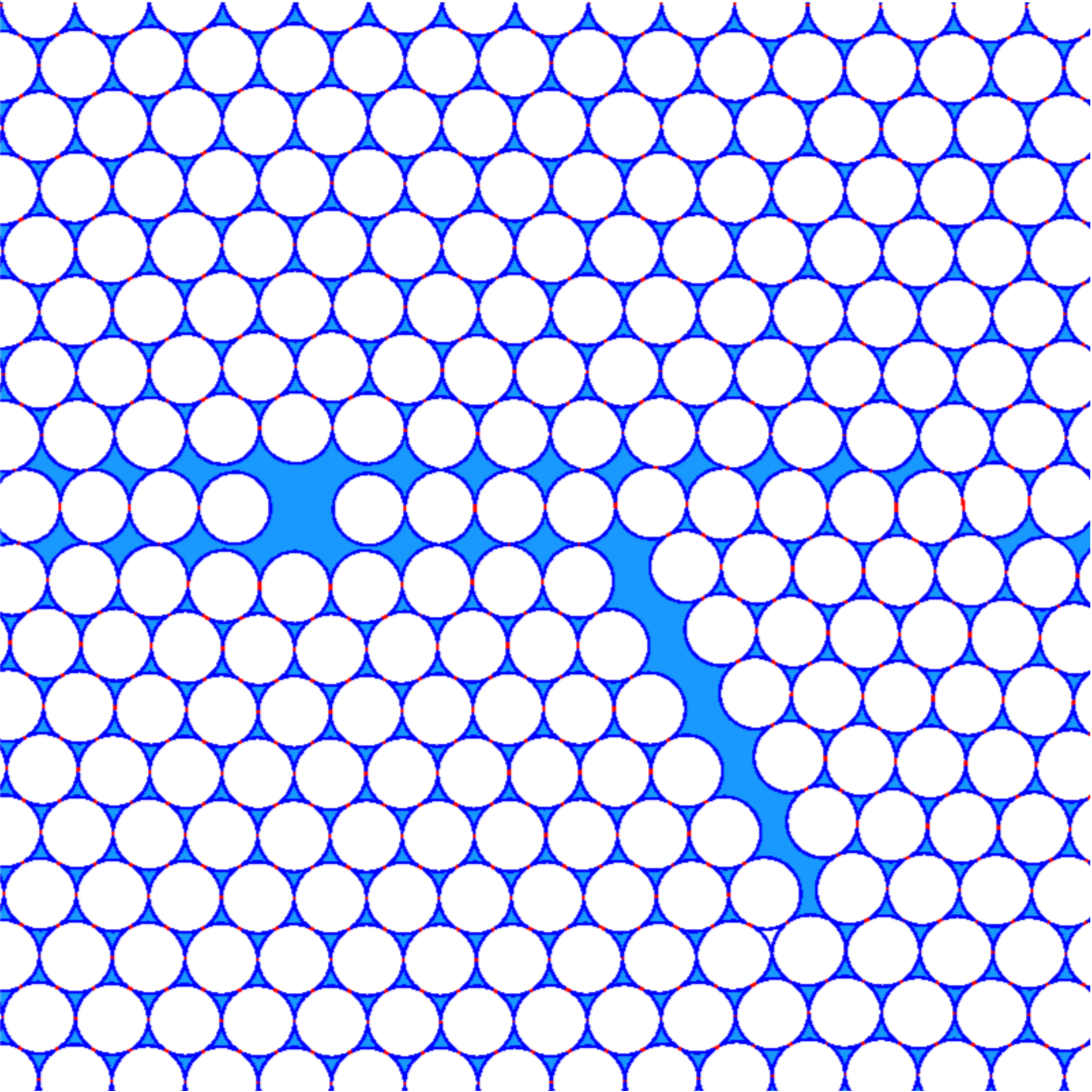}\\
(b) \includegraphics[width=0.7\textwidth]{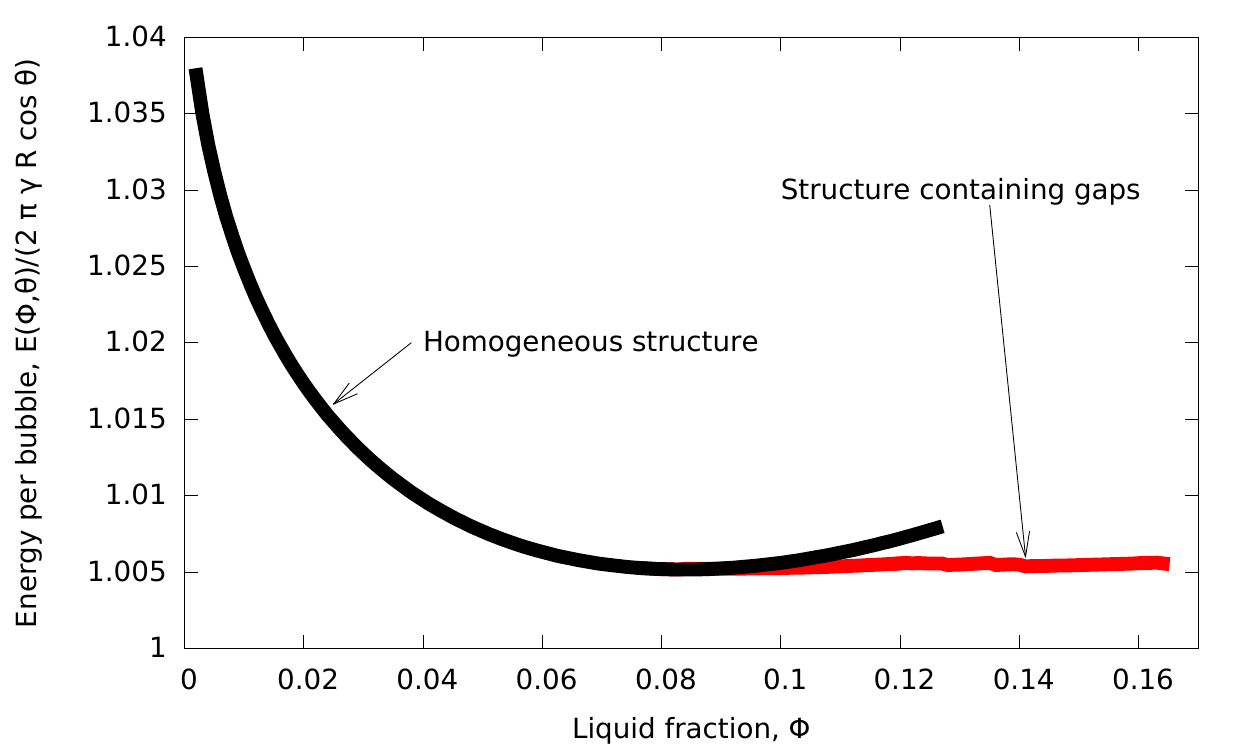}
\end{center}
\caption{
(a) Example of a Surface Evolver simulation of a monodisperse foam
($\phi=0.12$, contact angle $\theta = 6.26^\circ$, 256 bubbles)
which shows the formation of inhomogeneities  upon an increase
in liquid fraction above $\phi_0^{\theta}=0.084$.  
The appearance of inhomogeneities was triggered by a random
perturbation of the ordered structure.
(b) Upon a further increase in liquid fraction the energy remains roughly
constant. 
}
\label{f:honey_phase_separated}
\end{figure*}

The formation of inhomogeneities in the ordered monodisperse structure 
can be simulated by considering a representative area of foam 
that contains a large number of bubbles, 
and increasing the liquid fraction 
by expanding the system while keeping the bubble areas fixed.
An example of such an inhomogeneous structure is shown in figure
\ref{f:honey_phase_separated}. 
Finite contact angles were included by assigning different values for the
line tension in the bubble-bubble interfaces ($2 \gamma_f$) and the Plateau border sides ($\gamma$); 
the contact angle is given by eqn.~(\ref{e:contact_angle}). 
(For further details see \ref{a:se}.)

The inhomogeneity does not arise spontaneously in these simulations, 
but requires a perturbation at a liquid
fraction just above $\phi_0^{\theta}$.
This is achieved by randomly displacing
all of the Plateau border vertices by a small distance (less than a typical Plateau
border width); this allows the system to escape
from the metastable branch 
by undergoing topological transitions 
that are triggered when film lengths go to zero. 
These result in the formation of cracks, or liquid pools, as illustrated in
figure~\ref{f:honey_phase_separated}. 
The requirement of a perturbation to trigger instability
is a familiar feature of highly
symmetric structures that are locally stable, for which alternative
structures of lower energy are available.
Subsequently, increasing or decreasing the liquid fraction keeps the foam
at roughly constant energy. Small fluctuations in the energy occur because, as the liquid fraction changes in this {\em finite} sample of foam, there are short periods during which the energy increases elastically, followed by topological transitions that reduce the energy. As the sample size increases, such fluctuations become less marked.

This instability is reminiscent of the observations by 
Abd el Kadar and Earnshaw~\cite{AbdElKadarEarnshaw1997} in experiments with monodisperse 2D 
bubble rafts in an hexagonal 
confinement. 
About 25 minutes after foam formation, cracks appeared within the perfectly ordered structure, rapidly leading to the formation of a hole in the monolayer. No bubbles are lost in this process.
The authors attribute the development of this inhomogeneity to the movement and build-up of local stresses, for example due to small variations in bubble size. It is conceivable that the existence of a finite contact angle plays a further role, although the concept cannot be applied straightforwardly to bubble rafts, as opposed to 2D foam trapped between plates.

\section{Disordered foams with finite contact angle}

\begin{figure}
\begin{center}
\includegraphics[ width=1.0\textwidth]{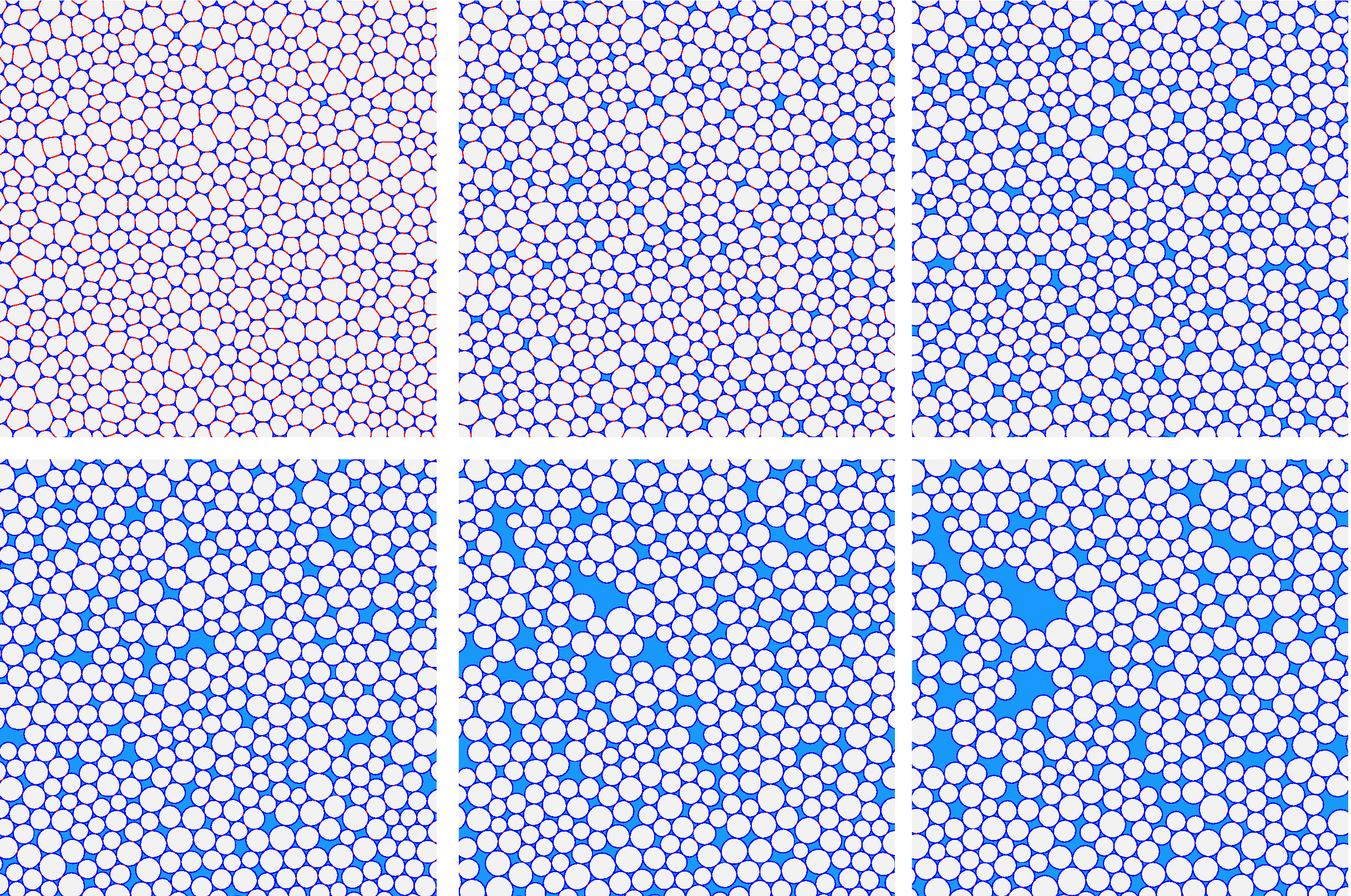}
\end{center}
\caption{
Surface Evolver simulation of a disordered foam of 1500 bubbles
and contact angle $\theta = 2.6^\circ$. An increase in liquid fraction leads
to the appearance of cracks or incipient flocculation. The images shown
correspond to close-ups of the foam at values of liquid fraction $\phi$
= 0.04, 0.08, 0.12, 0.16, 0.20 and 0.24 (from top left to bottom right).
}
\label{f:crack}
\end{figure}

Having established the effect of a finite contact angle for ordered foams
we now turn to the results of Surface Evolver simulations of
{\em disordered} foams, as exemplified in figure \ref{f:crack}.
The results presented below are based on five samples with $N=1500$ bubbles
and different values of the contact angle $\theta$ 
between $2.6^\circ$ and $15.9^\circ$.
The foams are polydisperse and
the variation in bubble areas differs between samples. 
Details of the simulations are presented in \ref{a:se}.

We again find that inhomogeneities develop as the liquid fraction $\phi$ is increased. 
However, these inhomogeneities do not necessarily result from metastable structures 
of the kind discussed in the previous section.
The disordered samples {\em spontaneously} undergo discrete topological transitions 
as the liquid fraction is increased; 
these occur when the length of a film between two neighbouring bubbles goes to zero, causing them to separate. 
In the simplest case, 
two bubbles that share a shrinking edge separate, 
and two three-sided Plateau borders merge to form a four-sided Plateau border, 
as shown in figure~\ref{f:t1sequence}. 
Another possibility occurs when non-adjacent edges 
on a Plateau border with four or more sides
come into contact and a new edge is formed. 
Various combinations of these topological transitions 
are responsible for the highly irregular liquid regions 
shown in figure~\ref{f:crack}.

\begin{figure}
\begin{center}
\includegraphics[ width=0.7\textwidth]{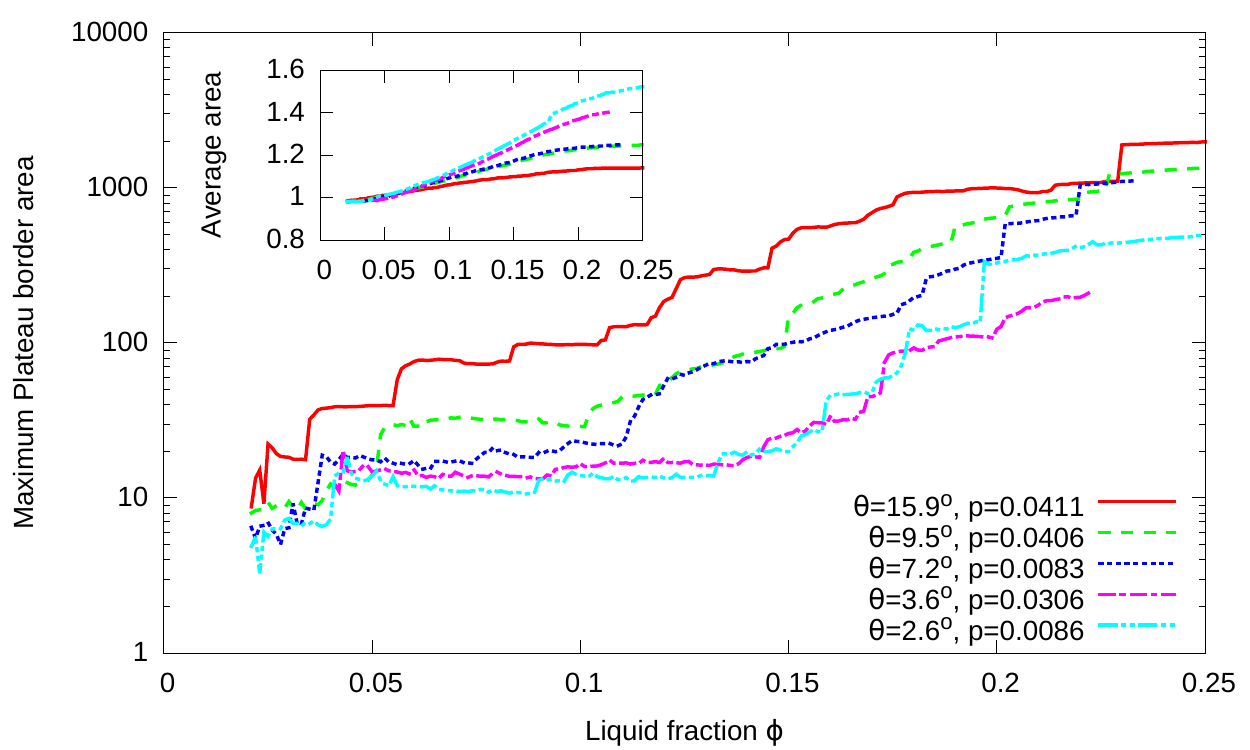}
\end{center}
\caption{
The development of inhomogeneities in disordered foams with finite contact
angle is illustrated by the observation that, while the average area of a
Plateau border $\langle A_{\rm pb} \rangle$ grows sublinearly with liquid fraction $\phi$ (inset), 
the area of the largest
Plateau border $A_{\rm pb}^{\rm max}$ (at each value of $\phi$) grows approximately exponentially (note the logarithmic vertical axis).
The data shown is for five simulations of disordered foams, with values of contact angle as indicated. 
The areas are normalized by the average area $\langle A_{\rm pb} \rangle$ that would result if Plateau borders did not merge, eqn.~(\ref{eq:pred_area}).
Note that the drops in area at low liquid fraction correspond to the splitting of a Plateau border with more than three sides into three-sided Plateau borders, as explained in \ref{a:se}.
}
\label{f:pbgrowth}
\end{figure}

Figure~\ref{f:pbgrowth} illustrates the growth of inhomogeneity as the liquid fraction is increased.
The average area of the Plateau borders $\langle A_{\rm pb} \rangle$ is given by
\begin{equation}
\langle A_{\rm pb} \rangle = 
\left( \frac{2 N}{N_{\rm pb}} \right) \frac{\langle A \rangle}{2} \frac{\phi}{1-\phi} ,
\label{eq:pred_area}
\end{equation}
where $N$ is the constant number of bubbles, 
$\langle A \rangle$ is the constant average bubble area, 
and $N_{\rm pb}$ is the number of Plateau borders, 
which decreases when they merge. 
If the Plateau borders do not merge, i.e. if they all remain three-sided, the pre-factor $2 N/N_{\rm pb}$ is unity.
The average Plateau border area, 
scaled by the value of $\langle A_{\rm pb} \rangle$ that would result if they didn't merge, 
is shown in the inset to figure~\ref{f:pbgrowth}. 
Note that the average Plateau border area 
rises more quickly at lower contact angles; 
this occurs because the tendency to combine is more pronounced 
when the contact angle is smaller, so the film lengths are smaller, 
and consequently, topological transitions are more frequent. 
The evolving inhomogeneity exhibits an acceleration in the growth 
of a few, increasingly larger, Plateau borders, while most remain small. 
The area of the largest Plateau border, $A_{\rm pb}^{\rm max}$, 
within a sample (figure \ref{f:pbgrowth}) increases roughly 
exponentially with liquid fraction.

\begin{figure}
\begin{center}
(a)\includegraphics[ width=0.7\textwidth]{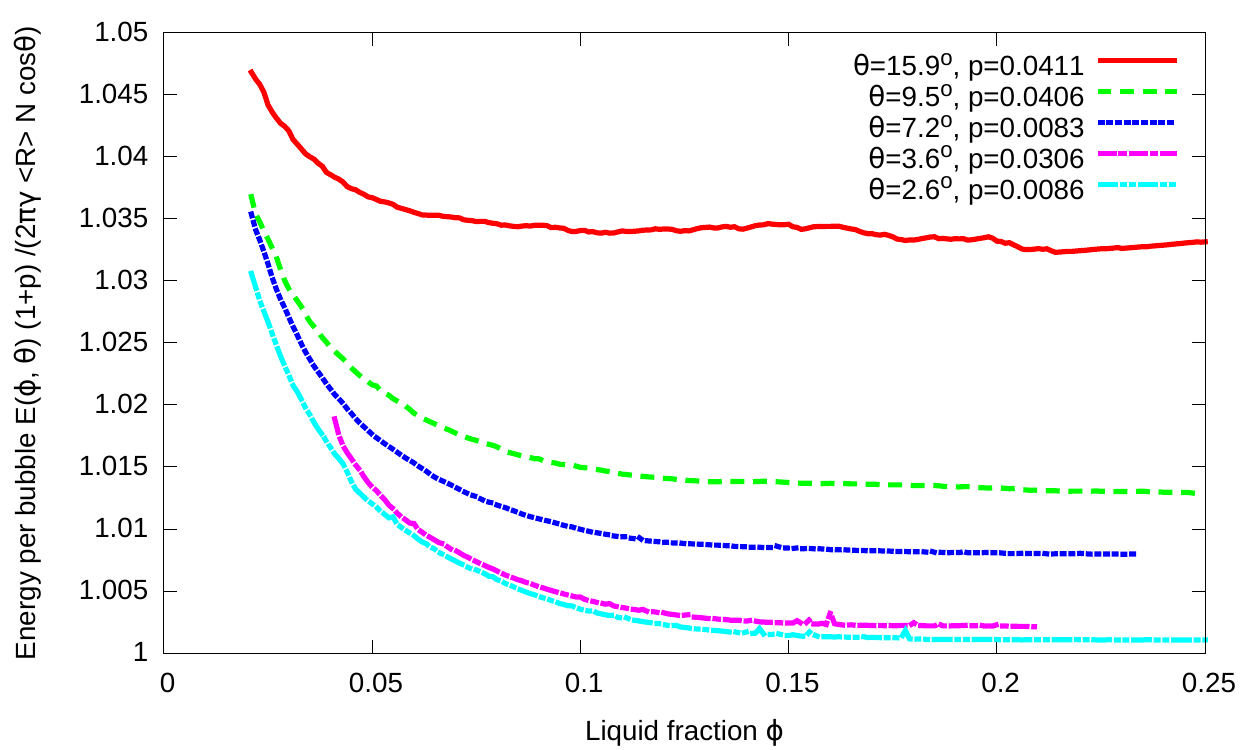}
\end{center}
\caption{
Variation of energy with liquid fraction $\phi$ for five disordered foams with
finite contact angles. The gradual
development of inhomogeneities, as shown in figure \ref{f:crack}, is accompanied by
a decrease in energy as a function of $\phi$ followed by saturation to a roughly constant value.  
The energy has been rescaled using the polydispersity
parameter $p$ defined in eqn.~(\ref{e:poly-p}). 
}
\label{f:se-energy}
\end{figure}

The energy $E(\phi,\theta)$ per bubble, plotted as a function of liquid fraction in figure~\ref{f:se-energy}, 
has been rescaled by $(1+p)$ to compensate for the polydispersity of the samples; 
$p$ is a measure of polydispersity introduced in~\cite{Kraynik03} and defined in \ref{a:2de}.
As in the case of ordered foams, 
an increase in contact angle leads to an increase in energy for a given value of the liquid fraction. 
For fixed contact angle, the energy decreases with increasing liquid fraction, 
at first steeply, and then more slowly. 
The initially monotonic decrease in energy with $\phi$ is consistent 
with the gradual development of inhomogeneities that we observe in disordered foams.

What should we expect the eventual state of a foam to be, 
were this wetting process to be continued to ever higher values of liquid fraction, 
beyond the usual 2D wet foam limit? 
We expect that the presence of a finite contact angle, 
which causes an effective adhesion between the bubbles, 
will lead to one large cluster of bubbles surrounded by liquid. 
This is again reminiscent of the clustering seen in 2D bubble rafts~\cite{Dennin}.

\section{Summary and conclusions}

Due to their increased stability, adhesive or ``sticky'' 3D emulsions, featuring droplet 
flocculation~\cite{bibette1993structure,aveyard1999flocculation,aveyard2002flocculation},
find many applications in the food and cosmetics
industries. Here we have discussed flocculation in the case of 
the analogous system of liquid foams.

Unlike the familiar case of a foam with zero contact angle, an ideal two-dimensional foam  with a finite contact angle develops an inhomogeneity for high liquid fraction $\phi$.
In an ordered foam this inhomogeneity appears at a critical value of $\phi$, but in a disordered foam there is a gradual development of inhomogeneity with increasing $\phi$. 
This implies that the notion of the wet limit is ill-defined in foams with finite contact angle.

As in other aspects of foam physics, our results for 2D foam should give
a general indication of the corresponding properties of 3D foams and emulsions. 
It is surprising that the subject of finite contact angles
has remained underdeveloped, and it is hoped that the present results will stimulate
further experiments and simulations. Indeed, there does not appear to be any 
2D data that could be compared to our findings.

In light of the described inhomogeneous structures, 
it may now be of interest to revisit earlier work on the ideal disordered 2D foam model, 
but including a finite contact angle. 
For example, does a finite contact angle affect the statistics of bubble rearrangements, 
perhaps by suppressing system-wide avalanches of topological changes, 
following  a small increase in $\phi$~\cite{DunneEtal2017}?

\section*{Acknowledgements}
SJC wishes to thank K. Brakke for advice about Surface Evolver
simulations and R. H\"ohler for useful discussions. SH and DW acknowledge the support 
of the MPNS COST Actions MP1106 ‘Smart and green interfaces’ and MP1305 ‘Flowing matter’.
The visit of AMK to the TCD Foams and Complex Systems group was
funded by the TCD Visiting Professors Fund and the Dublin Graduate Physics Program, and his visit to Aberystwyth was funded by the Distinguished Visiting Fellowship Scheme of the Royal Academy of Engineering.
Financial support from Science Foundation Ireland (SFI) under grant number 13/IA/1926 (SH, DW) and the 
European Space Agency ESA, Project microG-Foam, AO–99–075 and 
contract 4000115113, `Soft Matter Dynamics' (SJC, SH, DW) is also gratefully acknowledged.

\appendix
\section{Simulations using the Surface Evolver}
\label{a:se}

We carried out simulations of both ordered (hexagonal) and disordered
(polydisperse) foams with finite contact angles using Ken Brakke's Surface Evolver
software~\cite{Brakke92}.

In principle, all the properties of an ordered hexagonal foam can be captured
from a single cell, considered with periodic boundary conditions. However,
since we seek an instability of this structure to an inhomogeneous state, we must
instead simulate many cells. We chose to reproduce the basic hexagonal
cell 256 times to form a 16x16 hexagonal lattice with periodic
boundary conditions.

The disordered foams are made as dry foams, also with periodic boundary conditions, from a
Voronoi construction in the usual way~\cite{Brakke2006};
these have $N=1500$ bubbles,
with average area $A$ close to 1, and different polydispersity.

Both ordered and disordered foams are turned into wet foams 
of liquid fraction $\phi \approx 0.03$ by adding small triangular Plateau borders at each 
three-fold vertex; the liquid fraction is set by the total area of 
all Plateau borders. They have no individual area constraints, 
and therefore all have the same
pressure.  

Each side of each Plateau border is associated 
with a fixed bulk interfacial 
tension $\gamma\ge \frac{1}{2}$; the interfacial tension in the thin liquid films is set to
$2\gamma_f = 1$.
The contact angle is then given by eqn.~(\ref{e:contact_angle}).
The current version of the Surface Evolver software does not allow for the
simulation of 2D foams with zero contact angle and we find that $\theta =
2.6^\circ$ is the minimum value that we can choose.
Examination of the
consequences of this limitation provided some initial motivation for this study.

In the Surface Evolver, each
edge is represented as a circular arc, and a local minimum of the
interfacial
energy is sought using up to $2\times 10^4$ iterations to achieve a
relative accuracy close to $10^{-6}$.

The liquid fraction is changed by increasing the area of liquid and
keeping the area of gas constant (so the size of the periodic box
increases). Small increments in $\phi$ allows us to explore
a large range of liquid fractions: at each step, the liquid fraction is 
increased by 0.001, up to about $\phi = 0.25$.
We used the gradient descent method for energy minimisation, with occasional
Hessian iterations, to move towards a
minimum of interfacial energy  $\sum_j \gamma_j L_j$. Here $L_j$  denotes the lengths of the
interfaces, and $\gamma_j$ is either $\gamma$ or $2\gamma_f$, depending on
whether the interface is associated with a Plateau border or a film. 

\begin{figure*}
\begin{center}
\includegraphics[width=0.7\textwidth]{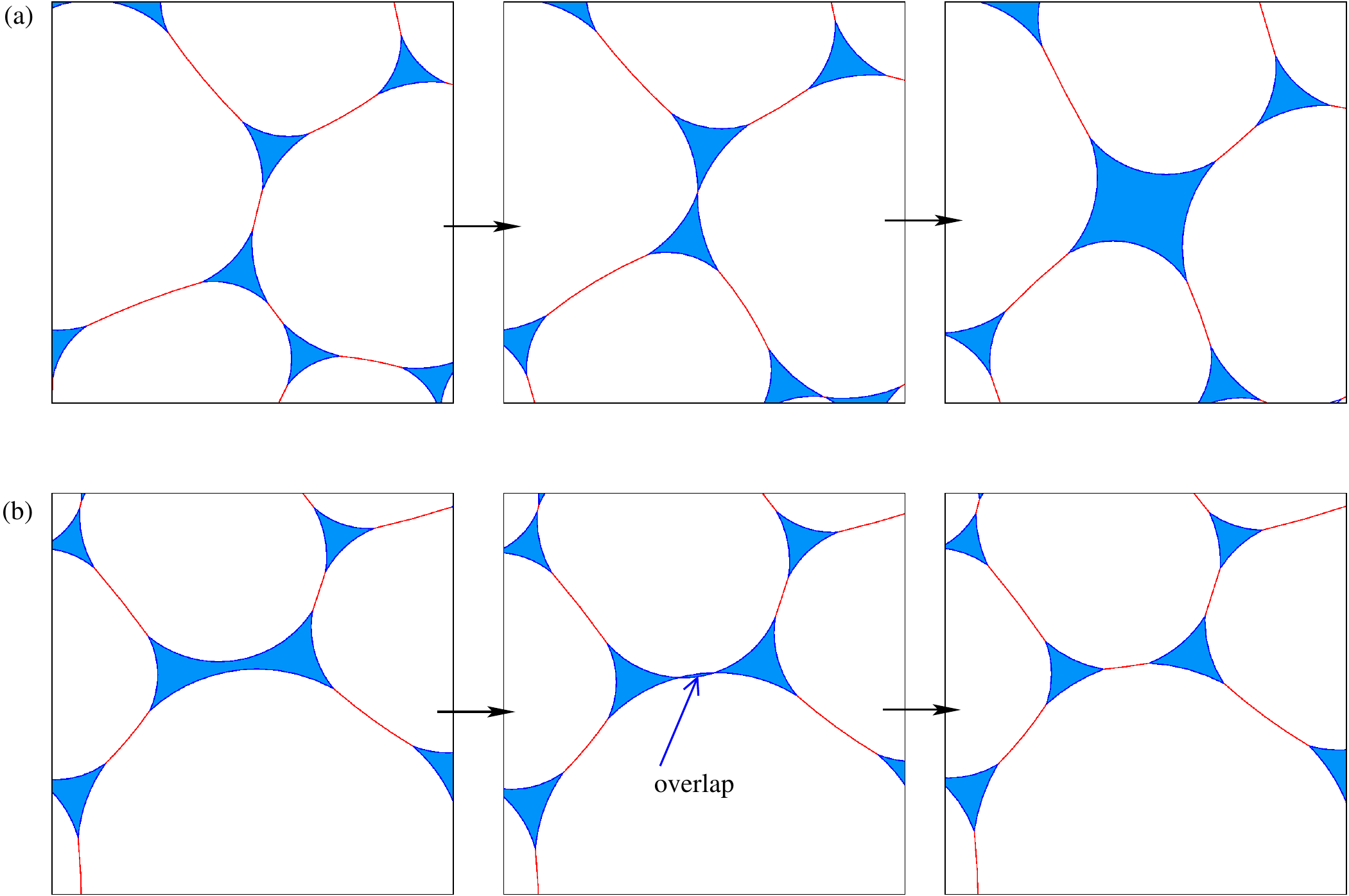}
\end{center}
\caption{
Numerical method for performing topological changes. (a) Two three-sided Plateau borders join when the interface between them shrinks beow a critical length. The four-sided Plateau border is created by deleting this short interface and merging the Plateau borders. (b) Two sides of a four-sided Plateau border approach each other; when they overlap, a new interface is inserted, creating two three-sided Plateau borders. A similar process is used for Plateau borders with more than four sides. Note that all Plateau borders have the same pressure, irrespective of their number of sides, and that their total area is conserved (rather than individual areas), so that no repartition of the liquid is required.
}
\label{f:t1sequence}
\end{figure*}

As the foam evolves, bubble rearrangements (topological changes) occur. 
We chose a critical film length of
$10^{-4}$ below which these are triggered. The difficult step is in
recognising when a four-sided Plateau border should split into two
three-sided Plateau borders 
(or, in general, a many-sided Plateau border 
should split into two
parts). To achieve this we check whether two sides of any Plateau border with more than three sides 
overlap, as illustrated in figure \ref{f:t1sequence}.
If they do, these two
sides are joined, to split the Plateau border. We briefly increase $\gamma$
and perform a few iterations, and then the minimization continues.

\section{Energy of a polydisperse 2D foam}
\label{a:2de}

The energy of a foam decreases with increasing polydispersity in the
bubble areas at fixed liquid fraction~\cite{Kraynik03}. We capture this dependence as follows.
Using the equivalent circular radius
$R$ of a bubble of area $A=\pi R^2$, we define polydispersity
to be 
\begin{equation}
p = \frac{R_{21}}{{\langle R^2 \rangle}^{1/2}} - 1. 
\label{e:poly-p}
\end{equation}
Here $R_{21}= \displaystyle
\frac{\langle R^2 \rangle}{\langle R \rangle}$ is the
Sauter mean radius in 2D and  the average, denoted with $\langle \rangle$,
is over all bubbles in the foam. 
The energy of an equilibrium, dry (i.e.,
$\phi = 0$), 2D foam can then be written as 
\begin{equation}
E = \frac{c}{1+p}
\frac{\gamma}{{\langle R^2 \rangle}^{1/2}},
\label{e:1pp}
\end{equation}
where the parameter $c$ depends very
weakly on bubble shape~\cite{KraynikEufoam2012}.
In plotting figure \ref{f:se-energy} we have assumed that the same scaling with $p$ applies at finite values of the liquid fraction $\phi$.

\end{document}